\documentclass[twocolumn,showpacs,superscriptaddress,preprintnumbers,amsmath,amssymb,prc]{revtex4-1}

\usepackage{graphicx}
\usepackage{dcolumn}
\usepackage{bm}
\usepackage{ifpdf}
\usepackage{epstopdf}
\usepackage{slashed}
\usepackage{amsfonts}
\usepackage{mathrsfs}
\usepackage{amssymb}
\usepackage{amsmath}
\usepackage{multirow}
\usepackage{CJK}
\usepackage{xcolor}
\usepackage{subfigure,dcolumn}
\usepackage[T2A,T1]{fontenc}
\usepackage[russian,english]{babel}
\usepackage{amsmath}
\usepackage{listings}
\usepackage{booktabs}
\usepackage{youngtab}
\usepackage[colorlinks,
linkcolor=blue,
anchorcolor=blue,
urlcolor=black,
citecolor=blue]{hyperref}

\begin{document}

\title{
Effects of the centrality determination method for the equation of state and nucleonic observables from Au+Au collisions at $\sqrt{s_{NN}}$ = 2.4 GeV}

\author{Xiaoqing Yue}
\affiliation{State Key Laboratory of Quantum Optics Technologies and Devices, Shanxi University, Taiyuan 030006, China}
\affiliation{School of Science, Huzhou University, Huzhou 313000, China}
\author{Pengcheng Li}
\affiliation{School of Science, Huzhou University, Huzhou 313000, China}
\author{Yongjia Wang}
\affiliation{School of Science, Huzhou University, Huzhou 313000, China}
\author{Qingfeng Li}
\email[Corresponding author, ]{liqf@zjhu.edu.cn}
\affiliation{School of Science, Huzhou University, Huzhou 313000, China}
\author{Fuhu Liu}
\affiliation{State Key Laboratory of Quantum Optics Technologies and Devices, Shanxi University, Taiyuan 030006, China}

\date{\today}

\begin{abstract}

Centrality determination remains one of the major sources of systematic uncertainty in intermediate-energy heavy-ion collision analyses, especially for probing the nuclear equation of state (EoS) at supra-saturation densities. To quantitatively assess the uncertainties associated with different centrality determination methods and to investigate their effects on final-state EoS-sensitive observables. Within the ultra-relativistic quantum molecular dynamics (UrQMD) model, Au+Au collisions at $\sqrt{s_{NN}}$=2.4 GeV are performed within a soft and a hard EoS. Event centrality is determined using the multiplicity of all charged particles ($M_\mathrm{ch}$) and two impact parameter-based centrality filters, one based on a geometrical interpretation and the other based on the Glauber Monte Carlo (MC) model, denoted as $b_{f}$ and $b_{r}$, respectively. It is shown that there exist significant differences between the real impact parameter distributions of event samples selected by $M_\mathrm{ch}$, $b_{f}$, and $b_{r}$, particularly between $M_\mathrm{ch}$ and $b_{r}$. When the $b_{f}$ is employed, uncertainties associated with centrality selection have a weaker influence on observables than the effects induced by the EoS. In contrast, when the $b_{r}$ is used, the influence of centrality-related uncertainties becomes more pronounced than that of the EoS. These results demonstrate that a rigorous and consistent mapping between $M_\mathrm{ch}$ and impact parameter is essential to impose quantitative constraints on the high-density nuclear EoS. Furthermore, our study indicates that the geometrical interpretation of centrality remains valid and consistent with dynamical multiplicity selection, whereas the Glauber MC-based centrality determination becomes unreliable at the investigated energy.

\end{abstract}
\maketitle
\section{Introduction}
Heavy-ion collisions (HICs) create extreme conditions of high energy and baryon density, under which nuclear matter can be compressed with several times saturation density ($\rho_{0}$), and offer a unique opportunity to investigate the properties of hot and dense nuclear matter \cite{Danielewicz:2002pu, Gyulassy:2004zy, Fukushima:2010bq, Andronic:2017pug}. 
High excitation and strong interaction in HICs lead to the production of a large number of particles, while the volume of nuclear matter rapidly expands \cite{Frankfurt:1981mk, Jacobs:2004qv}. 
The analysis of the information of the final-state observed particles can provide key insights into the properties of dense nuclear matter and impose critical constraints on its equation of state (EoS). 

In HICs, the spacetime structure of the particle emission source and the final state observables are usually closely related to the centrality, which can be quantified by the impact parameter $b$, defined as the perpendicular distance between the center-of-mass of the projectile and target.  
In theoretical simulations of HICs, the impact parameter serves as the input quantity. 
Nevertheless, in the real HIC experiments, the centrality cannot be directly detected and is generally determined by the final sensitive observables which can be detected but with the help of the predictions of a theoretical model, such as the total number of hits ($N_{\text{hit}}$) \cite{HADES:2017def,STAR:2022etb}, the number of reconstructed tracks ($N_{\text{trk}}$) \cite{Zhang:2017xda, STAR:2022etb}, the multiplicity of all charged particles ($M_\mathrm{ch}$) \cite{Cavata:1990gk,Kim:1992zza}, and the ratio of total transverse to longitudinal kinetic energy in the center-of-mass system ($ERAT$) \cite{FOPI:1996cjz}. 
However, there does not exist a strictly one-to-one relationship between the impact parameter and these observables. 
In our previous work, it was found that the true impact parameter of events selected with the impact parameter filter always has a certain distribution width, which must have a considerable influence on the collective flow in HICs at beam energies below $0.15A$ GeV \cite{Li:2020lup}. 
Subsequently, we further developed the method to infer the impact parameter by using machine learning and applied it to the analysis of experimental data at Sn+Sn collisions at $0.27A$ GeV detected by the S$\pi$RIT Time Projection Chamber \cite{Li:2020qqn,Li:2021plq,Tsang:2021rku}. 
It should be noticed that obtaining the $b$ values or determining the centrality, especially for more central collisions, from HIC experiments is still challenging \cite{Mallick:2021wop}.

Although intermediate-energy HICs occur only on a time scale of several tens of fm/$c$, the production and evaluation of the particles are dominated by the strong interaction, and the understanding of the strong interaction and the properties of nuclear matter remains limited \cite{Sorensen:2023zkk}. 
To explore the properties of nuclear matter under a wide range of densities, one common method is to compare the HIC experimental data from terrestrial laboratories with theoretical results from transport model simulations \cite{TMEP:2022xjg}. 
The density-dependent nuclear EoS is an essential ingredient of the transport models and has a great effort on particle production and collective behaviour \cite{Fuchs:2000kp,TMEP:2022xjg,Roca-Maza:2018ujj, Sorensen:2023zkk}. 
In recent years, the EoS around $\rho_{0}$ has been much better constrained, while the EoS at 2-4 times $\rho_{0}$ has large uncertainties \cite{Huth:2021bsp,Tsang:2023vhh}. 
The relative $K^{+}$ multiplicities per nucleon in Au+Au and C+C collisions observed by the KaoS Collaboration favour a soft EoS \cite{Fuchs:2000kp}. 
Nevertheless, by adopting a hard EoS in the Jet AA Microscopic transport model, the first- and second-order collective flow of light nuclei measured by the STAR Collaboration can be qualitatively reproduced \cite{STAR:2021ozh}. 
Further, the two-pion interferometry at GeV energies favours a relatively stiff EoS up to 4$\rho_{0}$, which then turns into a soft EoS at high densities \cite{Li:2022iil}. 
Furthermore, in the recent Ref. \cite{Li:2025iqq}, it was pointed out that within the Bayesian statistical framework and the isospin-dependent Boltzmann-Uehling-Uhlenbeck transport model, the proton flow excitation function data from FOPI indicate a gradual hardening of EoS as its density and temperature increase in reactions with higher beam energies. 
In addition, for the neutron star, the larger masses and radii favour a stiff EoS, but the hyperons will soften the EoS and reduce the maximum mass of the neutron star \cite{Oertel:2016bki,Li:2024lmd,Ye:2024meg}.

To more accurately constrain the information of EoS in dense nuclear matter by comparing the experimentally measured observables with the results of transport model simulations, the effects of the centrality determination method on the final state observables should be paid much more attention right now. 
In this work, based on the Ultra-relativistic Quantum Molecular Dynamics (UrQMD) model, the hard-EoS ($K_{0}$ = 380 MeV) and soft-EoS ($K_{0}$ = 200 MeV) with momentum dependence (labeled as HM and SM) are used to simulate Au+Au collisions at a typical SIS18 energy $\sqrt{s_{NN}}$=2.4 GeV (corresponding to the gold beam with a kinetic energy of $E_{\rm{beam}}$=1.23$A$ GeV on a fixed target), 
and the final-state charged-particle multiplicity $M_\mathrm{ch}$ and two initial-input impact parameter-based methods are used to select events within a certain centrality. One is given by the definition of geometrical centrality, denoted as $b_{f}$. 
Another is taken from Ref.~\cite{HADES:2017def}, which is based on the Glauber Monte Carlo (MC) model, denoted as $b_{r}$. 
However, the applicability of the Glauber MC model at low energies, such as 2.4 GeV, is questionable because the model assumes that the particle multiplicity is proportional to the number of participants estimated by the Glauber model, an assumption that works well at higher energies \cite{OmanaKuttan:2023cno}. 
Therefore, it is important to test various centrality (impact parameter) determination methods, evaluate and understand the influence of both the EoS and the centrality determination method on particle yield, collective flow, and transverse momentum distribution with great care.

The paper is organized as follows: Firstly, we briefly introduce the UrQMD model version in use, the centrality determination methods, and the analyzed observables in Sec. \ref{Meth.}. Then, the effects of the EoS and the centrality determination method on final observables are shown and discussed in Sec. \ref{result}. Finally, a summary and outlook is given in Sec. \ref{summ}.

\section{Methodology}\label{Meth.}

\subsection{UrQMD model}
The UrQMD model is a microscopic many-body approach to simulate the reaction of $pp$, $pA$, $AA$ systems, where the hadrons are represented by Gaussian wave packets with finite width and propagated according to Hamilton's equations of motion \cite{Bleicher:2022kcu,Bass:1998ca,Bleicher:1999xi}. 
Previous studies have illustrated that the mean-field potential plays an essential role in describing HICs at low and intermediate energies \cite{Steinheimer:2018rnd,Li:2021sdc}. Thus, the mean-field mode of the UrQMD model is adopted in this work. 
In this mode, the Hamiltonian of hadrons consists of the kinetic energy and the effective interaction potential energy, and the latter includes the two-body and three-body Skyrme terms (so-called density-dependent terms), the momentum-dependent term, and the Coulomb term as its basic components \cite{Li:2011zzp,Wang:2020vwb}, which can be expressed as

\begin{equation}\label{Vsimp}
\begin{aligned}
U =  \alpha \left(\frac{\rho}{\rho_{0}}\right)+\beta\left(\frac{\rho}{\rho_{0}}\right)^{\gamma} + U_{\text{md}} + U_{\text{Coul.}},
\end{aligned}
\end{equation}
and
\begin{equation}\label{Vsimp}
\begin{aligned}
U_{\text{md}} = t_{\text{md}} \ln^2 \left[ 1 + a_{\text{md}} (\mathbf{p}_i^2 - \mathbf{p}_j^2) \right] \frac{\rho}{\rho_0}.
\end{aligned}
\end{equation}
Here $\alpha$, $\beta$, $\gamma$, $t_{\rm{md}}$ and $a_{\rm{md}}$ are parameters that determine the stiffness of the nuclear matter EoS, which can be reified by the incompressibility $K_{0}$ \cite{Danielewicz:2002pu}. 
For HICs at intermediate energies, the momentum-dependent effect has a stronger effect on the observables than that of the isospin effect \cite{Ye:2018vbc,Du:2023ype,Steinheimer:2024eha}.  
Therefore, the simplified potential energy mentioned above has been widely used to investigate the non-equilibrium dynamics of HICs at GeV energies, and also adopted in this work, the parameter sets used here are shown in Table \ref{Table.1}.

In addition, it was pointed out that the isospin effect in cluster recognition has a small impact on the collective flow of Au+Au collisions at $E_{\rm{beam}}$=0.15 to 0.4$A$ GeV in Ref. \cite{Wang:2013wca}. Thus, the default isospin-dependent minimum spanning tree algorithm of the UrQMD model is applied for cluster identification at 100 fm/$c$ in this work. In this algorithm, if the relative distances and momenta of two nucleons are smaller than $R_{0}$ = 2.8 fm (proton-proton) or 3.8 fm (proton-neutron or neutron-neutron) and $P_{0}$ = 0.25 GeV/$c$, respectively, they are considered to belong to the same fragment \cite{Zhang:2012qm,Li:2016wkb,Kireyeu:2022qmv}. 
And by the preliminary test, it is found that the isospin effect of the coalescence and the stopping time of the simulation do not threaten the following conclusions.  

\begin{table}[t]
    \centering
    \caption{Parameter sets of the nuclear equation of state.}
    \label{Table.1}
    \vspace{2mm}
    \renewcommand{\arraystretch}{1.5}
    \begin{tabular}{l c c c c c c}
    \hline\hline
    EoS & $\alpha$(MeV) & $\beta$(MeV) & $\gamma$ & $t_{md}$(MeV) & $a_{md}$($\frac{c^2}{\text{GeV}^2}$) & $K_{0}$(MeV)\\
    \midrule
    SM  &  -393 & 320 & 1.14 & 1.57 & 500 & 200 \\
    HM  &  -138 & 60 & 2.084 & 1.57 & 500 & 380 \\
    \hline\hline
    \end{tabular}
\end{table}

\subsection{Centrality determination}

\begin{figure}[t]
    \centering
    \includegraphics[width=\linewidth]{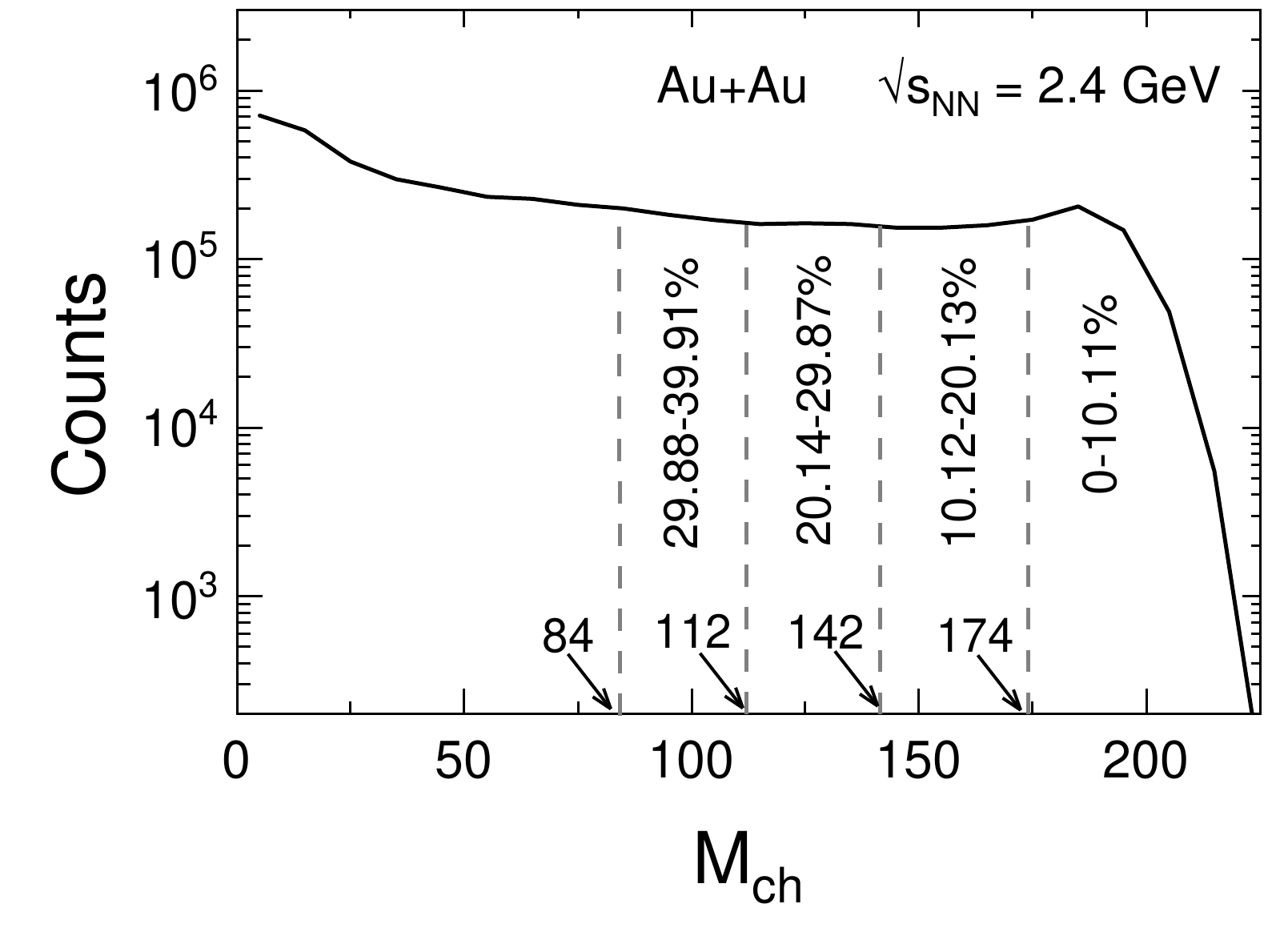}
    \caption{(Color online) The distribution of the multiplicity of all charged particles $M_\mathrm{ch}$ from Au+Au collisions at $\sqrt{s_{NN}}=2.4$ GeV with $b$=0-$b_{\text{max}}$.}
    \label{fig:1}
\end{figure}

In this work, 5 million Au+Au collisions at $\sqrt{s_{NN}}$=2.4 GeV with impact parameter up to $b_{\max}=1.15\times(A_{P}^{1/3}+A_{T}^{1/3})$ are simulated with $bdb$ weighted distribution. 
To estimate the centrality $\bar{c}_{\rm{est}}$, on the one hand, the total charged particle multiplicity $M_\mathrm{ch}$ is used \cite{Cavata:1990gk,Li:2017tom, Li:2020lup},

\begin{equation}\label{best0}
\begin{split}
	c_{\rm{est}}(M_{\text{ch}}\geqslant M_{\text{ch}}^{'})&= \frac{\sum_{M_{\text{ch}}^{'}}^{\infty}\sigma(M_{\text{ch}})}{\sum_{0}^{\infty}\sigma(M_{\text{ch}})}=\frac{\sum_{M_{\text{ch}}^{'}}^{\infty}N(M_{\text{ch}})}{\sum_{0}^{\infty}N(M_{\text{ch}})},
\end{split}
\end{equation}

\begin{equation}\label{cest}
	\bar{c}_{\text{est}}(M_{\text{ch}}\geqslant M_{\text{ch}}^{'})=\frac{1}{2}(c_{\text{est}}(M_{\text{ch}}^{'}+1)+c_{\text{est}}(M_{\text{ch}}^{'}-1)).
\end{equation}
Here, $\sigma(M_{\text{ch}})$ and $N(M_{\text{ch}})$ represent the cross section and the number of events with multiplicity equal to $M_{\text{ch}}$, respectively. 
For a given value of $M_{\text{ch}}^{'}$, the estimated centrality $\bar{c}_{\text{est}}$ with $M_{\text{ch}}\geqslant M_{\text{ch}}^{'}$ can be approximately determined by Eqs. \ref{best0} and \ref{cest}, as shown in Fig. \ref{fig:1}. 
On the other hand, the centrality can be deduced by the input impact parameter ($b_{f}$) by
\begin{equation}\label{b_c}
c=\left(\frac{b_{f}}{b_{\rm{max}}}\right)^2.
\end{equation}

In addition, due to the events selected by $M_\mathrm{ch}$ have a broader distribution of the true impact parameter \cite{Li:2017tom, Li:2020lup}. 
And in Ref. \cite{HADES:2017def}, the impact parameter ranges ($b_{r}$) corresponding to the various centralities are deduced through $N_{\rm{trk}}$ for the investigated system and energy. 
To make a meaningful comparison, the events with the same centrality, but filtered by $M_\mathrm{ch}$, $b_{f}$, and $b_{r}$ are analysized and compared. 
The centrality classes and the corresponding ranges of $M_\mathrm{ch}$, $b_{f}$, and $b_{r}$ are shown in Tab. \ref{Table.2}. It should be pointed out that due to the discreteness of $M_\mathrm{ch}$,  the cumulative multiplicity distribution to jump in integer step, the centrality boundaries determined via Eqs. \ref{best0} and \ref{cest} cannot always coincide exactly with round percentage values. Here, the closest possible value is used to define the corresponding centrality interval. 

\begin{table}[t]
    \centering
    \caption{Centrality classes and the corresponding ranges of $M_\mathrm{ch}$, $b_{f}$ and $b_{r}$.}
    \label{Table.2}
    \vspace{2mm}
    \renewcommand\arraystretch{1.5}
    \setlength{\tabcolsep}{6pt}
    \begin{tabular}{c c c c}
    \hline\hline
    Centrality & $M_\mathrm{ch}$ & $b_{f}$ (fm) & $b_{r}$ (fm) Ref. \cite{HADES:2017def} \\
    \midrule
    0-10.11\%   &  $\geq$ 174 & $\leq$ 4.26  & $\leq$ 4.70 (0-10\%) \\
    10.12-20.13\% &  173-142  & 4.27-6.00  & 4.71-6.60 (10-20\%) \\
    20.14-29.87\% &  141-112  & 6.01-7.31  & 6.61-8.10 (20-30\%) \\
    29.88-39.91\% &  111-84  & 7.32-8.45  & 8.11-9.30 (30-40\%) \\
    39.92-100\% & $\leq$ 83 &  $\geq$ 8.46  &  $\geq$ 9.31 (40-100\%) \\
    \hline\hline
    \end{tabular}
\end{table}

\begin{figure*}[t]
    \centering
    \includegraphics[width=1\linewidth]{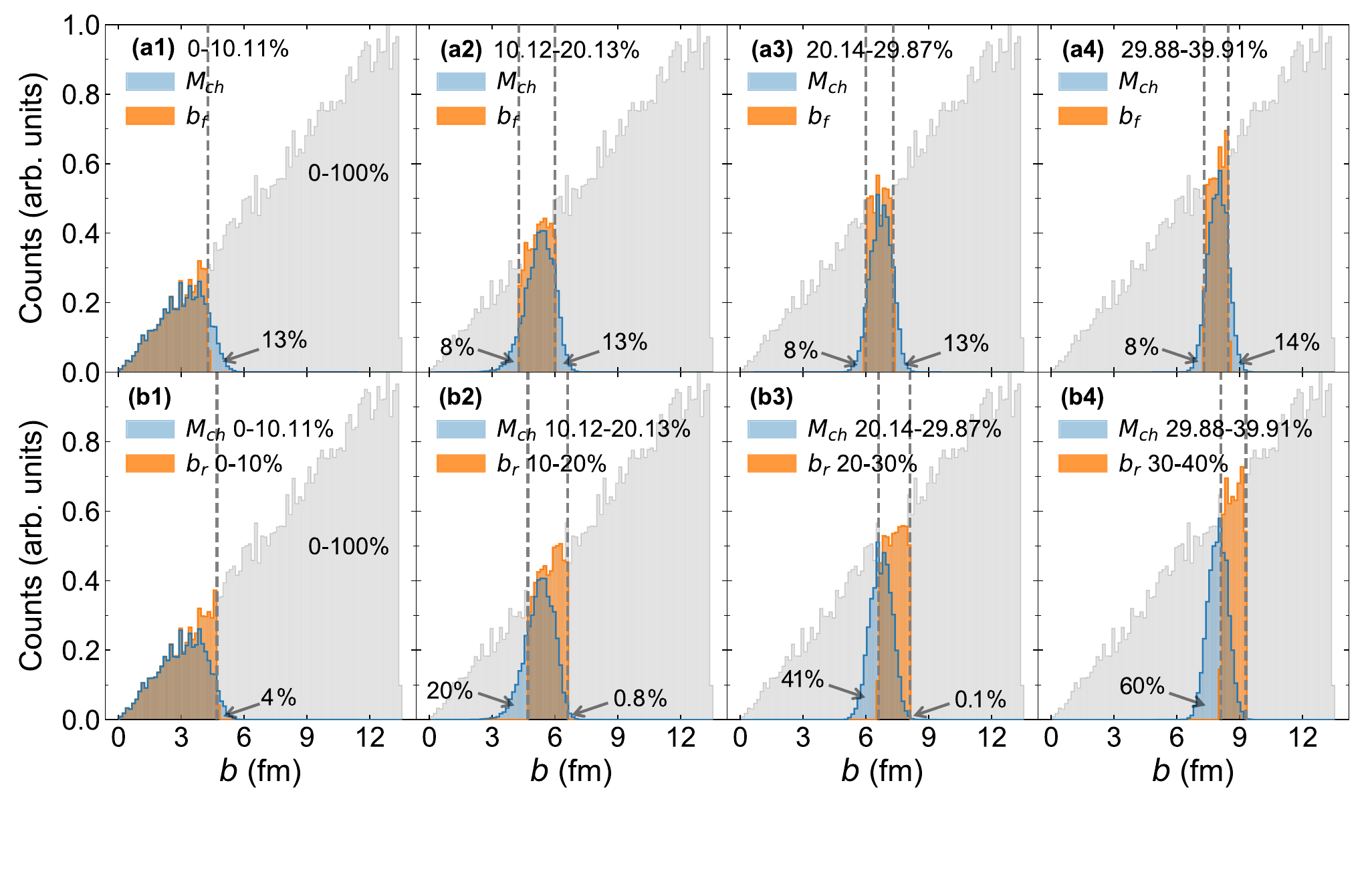}
    \caption{(Color online) The impact parameter distributions of the event samples selected by $b_{f}$, $b_{r}$, and $M_\mathrm{ch}$ given in Tab.~\ref{Table.2} from Au+Au collisions at $\sqrt{s_{NN}}$ = 2.4 GeV with $b$=0-$b_{\text{max}}$. }
    \label{fig:2}
\end{figure*}

Figure \ref{fig:2} shows the impact parameter distributions of the event samples selected by $M_\mathrm{ch}$, $b_{f}$, and $b_{r}$ given in Tab.~\ref{Table.2}. The top panels are for $M_\mathrm{ch}$ and $b_{f}$, while the bottom panels are for $M_\mathrm{ch}$ and $b_{r}$.
The bottom grey layers denote the distributions of events with impact parameter up to $b_{\text{max}}$, i.e. 0-100\% centrality. 
The pairs of panels [(a1), (b1)], [(a2), (b2)], [(a3), (b3)], and [(a4), (b4)] correspond to the 0-10\%, 10-20\%, 20-30\%, and 30-40\% centrality classes, respectively.
The middle orange layers represent the distributions of events selected by $b_{f}$ and $b_{r}$. 
While the top blue layers, with its Gaussian-like distribution, represent those of events selected by $M_\mathrm{ch}$. 
It can be found that for the same centrality, the impact parameter distributions of event samples selected by $M_\mathrm{ch}$ are wider than those of event samples selected by $b_{f}$ and $b_{r}$. 
For 0-10\% centrality, the events selected by $M_\mathrm{ch}$ have more events with larger impact parameter than the events selected by $b_{f}$, but fewer events with large impact parameters than the events selected by $b_{r}$. 
As for larger centrality, among the event samples selected by $M_\mathrm{ch}$, approximately 8\% of the events have impact parameters smaller than the lower bound of the $b_{f}$ range, while about 13\% have impact parameters larger than its upper bound at 10-20\% and 20-30\% centrality collisions. These deviations remain nearly constant with increasing centrality. 
However, the differences between the event samples selected by $b_{r}$ and $M_\mathrm{ch}$ amount to approximately 20\% and 0.8\% for the 10-20\% centrality class, 41\% and 0.1\% for 20-30\%, and 60\% and 0.06\% for 30-40\%. 
As the centrality increases, the fraction of events with larger $b$ in the samples selected by $b_{r}$ decreases, while that of events with smaller $b$ increases. 
With these difference in $b$ distribution between the event samples select by different centrality filters, the final state observables will be influenced by the centrality filters to some extend.

Additionaly, to assess the uncertainties associated with the potential impact of cuts, such as the rapidity and transverse momentum, and the definition of $b_{\rm{max}}$, the $0.3 < p_\mathrm{T} < 3.0$ GeV/c and $|y| \le 0.5$, and the $b_{\max}=1.12\times(A_{P}^{1/3}+A_{T}^{1/3})$ are adopted to have a test. 
It is found that the total $M_\mathrm{ch}$ is influenced, but the differences in the impact parameter distribution between the event samples selected by $M_\mathrm{ch}$ and $b_{f}$, as well as $b_{r}$ are similar, and the results shown in Sec. \ref{result} are not threatened. 

\subsection{Observables}
At the investigated energy of Au+Au collisions, the production of free protons dominates the charged particle yield, accounting for approximately 35\% in central collisions with $b$=0 fm, and can be detected with high efficiency. Thus, the yield and the spectra of free protons, such as the rapidity ($y$) distributions and transverse momentum ($p_\mathrm{T}=\sqrt{p_{x}^{2}+p_{y}^{2}}$) distributions, will contain the information of the whole dynamical evolution. By analysing the $y$ distribution of free protons and the closely related stopping observables, the deformation of the colliding nuclei \cite{JiaYe:2021lgg}, the stiffness of the EoS \cite{FOPI:2010xrt}, the in-medium nucleon-nucleon cross section \cite{Li:2018wpv}, and the sequential decay \cite{Xiao:2023pqs} are explored. 

Besides the rapidity distribution of free protons, the anisotropic collective flow of free protons is also a sensitive probe of the EoS, and can be obtained by Fourier expansion of the final azimuthal angle ($\phi$) distribution of the particle \cite{Stoecker:1986ci, Reichert:2023eev},
\begin{equation}
\frac{dN}{d\phi} = 1 + 2 \sum_{n=1}^{\infty} v_n \cos \left[ n(\phi - \Psi_{RP}) \right],
\end{equation}
where $v_{n}$ represents the $n^\mathrm{th}$ order harmonic flow coefficient, $\Psi_{RP}$ is the azimuthal angle of the reaction plane and is fixed at zero in the following analysis for simplicity. 
For $n$ = 1, $v_1 \equiv \langle \cos\phi \rangle = \left\langle \frac{p_x}{p_\mathrm{T}} \right\rangle= \left\langle \frac{p_x}{\sqrt{p_{x}^{2}+p_{y}^{2}}} \right\rangle$ is called directed flow, which comes from the deflection of the projectile (target) matter in the positive (negative) $x$ direction in non-central HICs, reflects the particle collective motion in the direction parallel to the reaction plane and the competition between the repulsive two-body collisions and the mean-field potential \cite{Reisdorf:1997fx,Herrmann:1999wu,Reichert:2024ayg}. 
For $n$ = 2, $v_2 \equiv \langle \cos(2\phi) \rangle = \left\langle \frac{p_{x}^{2}-p_{y}^{2}}{p_\mathrm{T}^2} \right\rangle$ is called elliptic flow, which originates from the pressure of the compressed region, is affected by the passage time of the spectator matter, characterizes the collective motion of the particles in the direction perpendicular to the reaction plane \cite{Snellings:2011sz}. 
In Refs. \cite{Gao:2022shr,Wang:2024ktk}, it was found that the $v_2$ in HICs at different energies is affected to varying degrees by the impact parameter, mean-field potential, and two-body collisions. 
Given the extensive accumulation of high-precision experimental data on $v_{1}$ and $v_{2}$ in intermediate-energy heavy-ion collisions, this paper focuses on $v_{1}$ and $v_{2}$ of free protons. 
For the higher-order flow coefficients, one can refer to Refs. \cite{Reisdorf:1997fx,Voloshin:2011mx,Heinz:2013th,HADES:2020lob} for detailed reviews of anisotropic collective flow in HICs. 

As described above, the $v_{1}$ and $v_{2}$ are closely related to the transverse momentum $p_\mathrm{T}$, and the $p_\mathrm{T}$ distribution of free protons is influenced by the EoS, thus, the detected free protons with various $p_\mathrm{T}$ experience different dynamical evolutions and carry the EoS information with different densities \cite{Wang:2024ktk}. 
In addition, the mean transverse momentum $\langle p_\mathrm{T} \rangle$ has been suggested to be influenced by the initial state conditions and the EoS in HICs \cite{Steinheimer:2022gqb, Tarasovicova:2024isp}, which can be expressed as
\begin{equation}
\langle p_\mathrm{T} \rangle=\frac{1}{N}\sum_{i=1}^{N}p_{T,i}=\frac{\int p_{T}\frac{dN}{dp_{T}}dp_{T}}{\int \frac{dN}{dp_{T}}dp_{T}}, 
\end{equation}
where $N$ is the total number of particles. 
Hence, the yield spectrum, collective flow, and (mean) transverse momentum will be systematically analyzed in the following to evaluate the effects of the centrality determination method and the EoS on the final state observables.

\section{Results and discussion}\label{result}

\begin{figure}[t]
    \centering
    \includegraphics[width=0.8\linewidth]{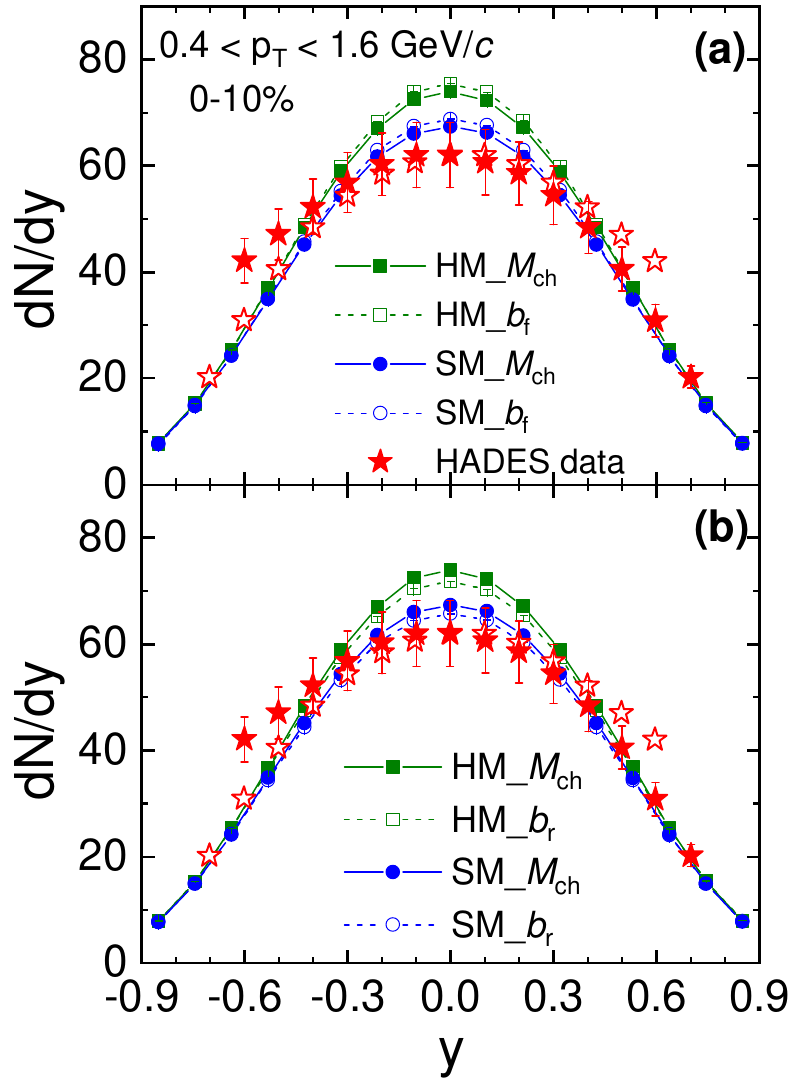}
    \caption{(Color online) Rapidity distribution of free protons within $0.4<p_\mathrm{T}<1.6$ GeV/$c$ in central (0-10\%) Au+Au collisions at $\sqrt{s_{NN}}$ = 2.4 GeV for different EoS and centrality determination methods. The experimental data (red stars) are taken from the HADES Collaboration \cite{HADES:dndy}. For the theoretical results, the error bars fall within the symbols.}
    \label{fig:3}
\end{figure}

\subsection{Free proton yield}

Firstly, Fig. \ref{fig:3} depicts the rapidity distribution of free protons in 0-10\% central Au+Au collisions at $\sqrt{s_{NN}}$ = 2.4 GeV with $0.4<p_\mathrm{T}<1.6$ GeV/c for different EoS and centrality determination methods. 
The results obtained with the SM EoS (blue circles) can better reproduce the experimental data \cite{HADES:dndy}, whereas those with the HM EoS (green squares) are slightly higher at mid-rapidity. 
At the investigated energy, a softer EoS allows stronger compression and higher densities, whereas a stiffer EoS leads to stronger repulsive pressure gradients and larger transverse flow during the high-density phase. 
This enhanced pressure redirects more nucleons toward mid-rapidity, resulting in a higher free-proton yield and a reduced cluster yield at mid-rapidity \cite{FOPI:2011aa,Ye:2018vbc}.   
Moreover, in the top panel, the results for events classified by $M_\mathrm{ch}$ (solid symbols) are lower than those for events selected by the $b_{f}$ (open symbols), while in the bottom panel, the $M_\mathrm{ch}$ classified results are higher than those selected by $b_{r}$. 
Since for the 0-10\% centrality, the events selected by $M_\mathrm{ch}$ have more events with larger impact parameter than the events selected by $b_{f}$, but fewer events with large impact parameters than the events selected by $b_{r}$, as shown in Fig.~\ref{fig:2}.

\begin{figure}[t]
    \centering
    \includegraphics[width=0.9\linewidth]{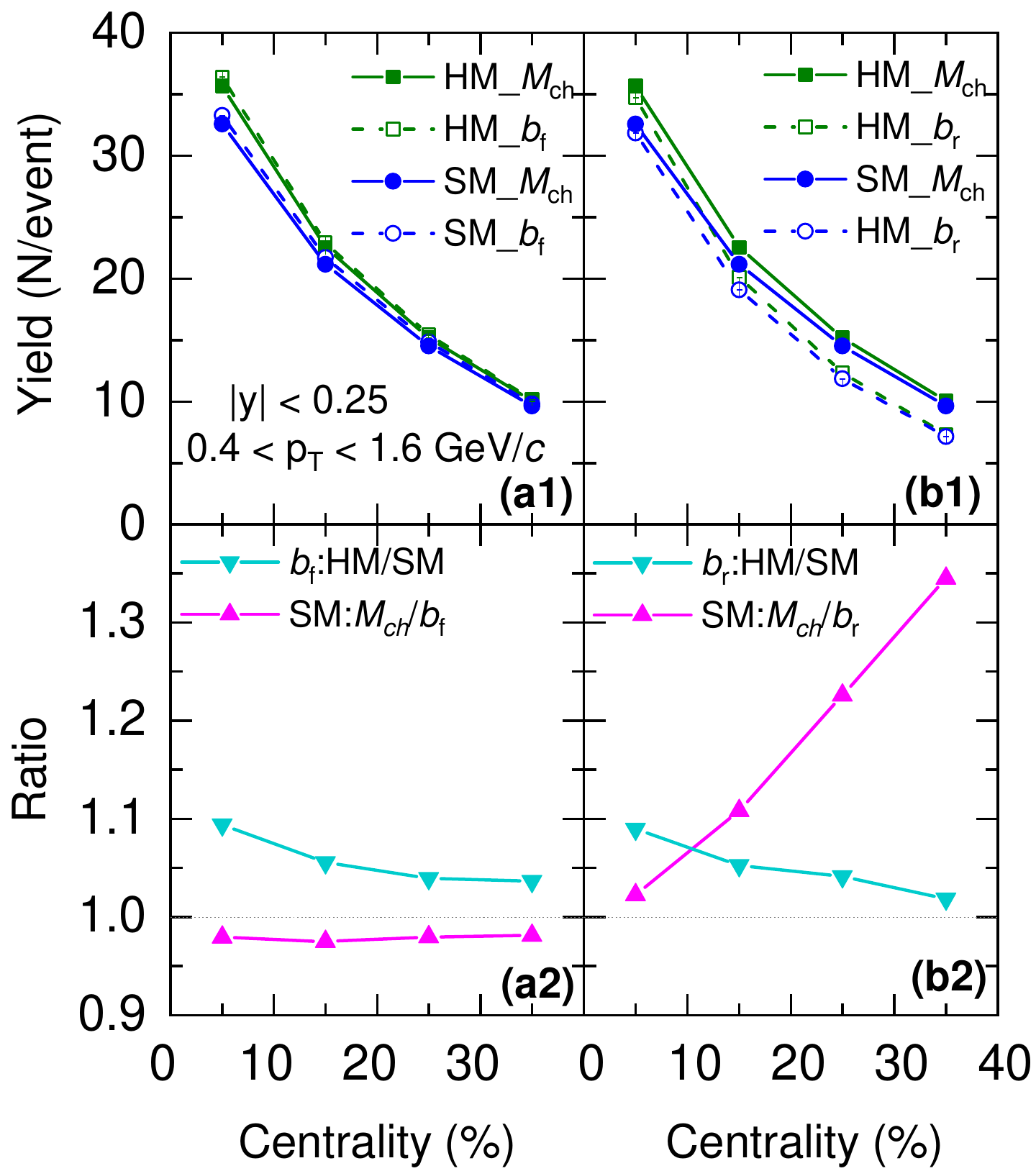}
    \caption{(Color online) Top panels: Centrality dependence of the free proton yields within $0.4<p_{\mathrm{T}}<1.6$ GeV/c and $|y|<0.25$ for four scenarios; symbol conventions follow Fig.~\ref{fig:3}.
    Bottom panels: Centrality dependence of the yield ratios between different EoS (cyan lines with downward triangles) and centrality determinations (magenta lines with upward triangles). 
    Where not visible, error bars are smaller than the symbols.}
    \label{fig:4}
\end{figure}

To further investigate the centrality dependence of the effect of EoS and centrality determination method on proton yields at mid-rapidity, the top panels of Fig. \ref{fig:4} show the proton yields at mid-rapidity as a function of the centrality within $0.4<p_{\mathrm{T}}<1.6$ GeV/c, the same line and symbol conventions are adopted as Fig. \ref{fig:3}. 
One can observe again that the free proton yields of simulations with a stiff EoS (green squares) are higher than those with a soft EoS (blue circles) for all $M_\mathrm{ch}$ (solid symbols), $b_{f}$ (open symbols), and $b_{r}$ (open symbols) scenarios. 
In panel (a1), the results of the events selected by $M_\mathrm{ch}$ and $b_f$ (i.e. the same color but different symbols) are almost close to each other for the four centrality classes. 
However, the difference between the results of events selected by $M_\mathrm{ch}$ and $b_{r}$ increases with centrality, as shown in panel (b1). 
In addition, the difference between HM and SM results (the same line type but different colors) decreases as centrality increases.

And, to clearly see the centrality dependence of these differences, the ratios in the free proton yield between different scenarios are calculated and shown in the bottom panels of Fig. \ref{fig:4}. 
Under the same centrality determination, the yield ratio between HM and SM ($b_{f}$: HM/SM, cyan lines with downward triangle) decreases slowly with increasing centrality, since with increasing centrality, the density of the system decreases, and the effects of the density-dependent potential interaction are reduced. 
However, the yield ratio between $M_\mathrm{ch}$ and $b_f$ (SM:$M_\mathrm{ch}/b_{f}$, magenta line with upward triangle in panel (a2)) under the same EoS remains basically unchanged, while that between $M_\mathrm{ch}$ and $b_{r}$ (SM:$M_\mathrm{ch}/b_{r}$, magenta line with upward triangle in panel (b2)) increases with increasing centrality significantly. 
These results indicate that the difference in the yields between the event samples selected by $M_\mathrm{ch}$ and $b_{f}$ is small, while that between the event samples selected by $M_\mathrm{ch}$ and $b_{r}$ is larger, and increases with increasing centrality.

From Fig. \ref{fig:2}, we have learned that although the event samples selected by $M_\mathrm{ch}$ have a wider impact parameter distribution than that of $b_{f}$ and $b_{r}$, 
for the event samples selected by $M_\mathrm{ch}$, there is a countervailing effect between the fraction of events with smaller and larger impact parameters beyond the $b_{f}$ bound for four centrality classes, which are unchanged with centrality increases.  Whereas the fraction of events with impact parameters smaller than the lower bound of $b_{r}$ increases and that with impact parameters larger than the upper bound of $b_{r}$ decreases with increasing centrality. 
Consequently, the results of events selected by $M_\mathrm{ch}$ are close to those of $b_{f}$, but the difference between $M_\mathrm{ch}$ and $b_{r}$ increases as centrality increases.

\subsection{Directed and elliptic flow}

\begin{figure}[t]
    \centering
    \includegraphics[width=0.95\linewidth, height=\textheight, keepaspectratio]{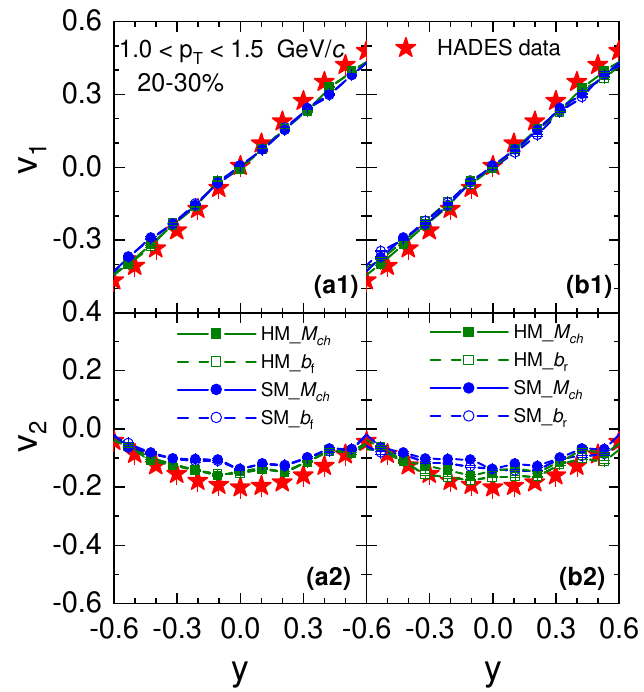}
    \caption{(Color online) Rapidity distribution of the directed flow $v_{1}$ [top panels (a1) and (b1)] and the elliptic flow $v_{2}$ [bottom panels (a2) and (b2)] of free protons at 20-30\% centrality with different impact parameter filters and EoS, respectively. The solid stars represent the experimental data taken from Ref. \cite{HADES:2020lob}. }
    \label{fig:5}
\end{figure}

Then, the effects of the centrality determination method on the collective flow under different EoS are explored. 
Figure \ref{fig:5} shows the directed flow $v_{1}$ (top panels) and elliptic flow $v_{2}$ (bottom panels) of free protons as a function of rapidity for 20-30\% centrality Au+Au collisions within $1.0<p_{\mathrm{T}}<1.5$ GeV/c. 
The symbol sets for the results from simulations with different EoS and centrality determination methods are the same as in Fig. \ref{fig:3}, and the experimental data are taken from Ref. \cite{HADES:2020lob}. 
For the $v_{1}$, the results for both scenarios appear to be close to each other and generally consistent with the experimental data within error bars, especially at the mid-rapidity.
As for the $v_{2}$, the absolute value of $v_{2}$ of HM is larger than that of SM, and the results of events selected by $b_{r}$ are slightly larger than those of $M_\mathrm{ch}$, while the results of events selected by $b_{f}$ and $M_\mathrm{ch}$ are nearly overlapping.

\begin{figure}[t]
    \centering
    \includegraphics[width=0.92\linewidth, height=\textheight, keepaspectratio]{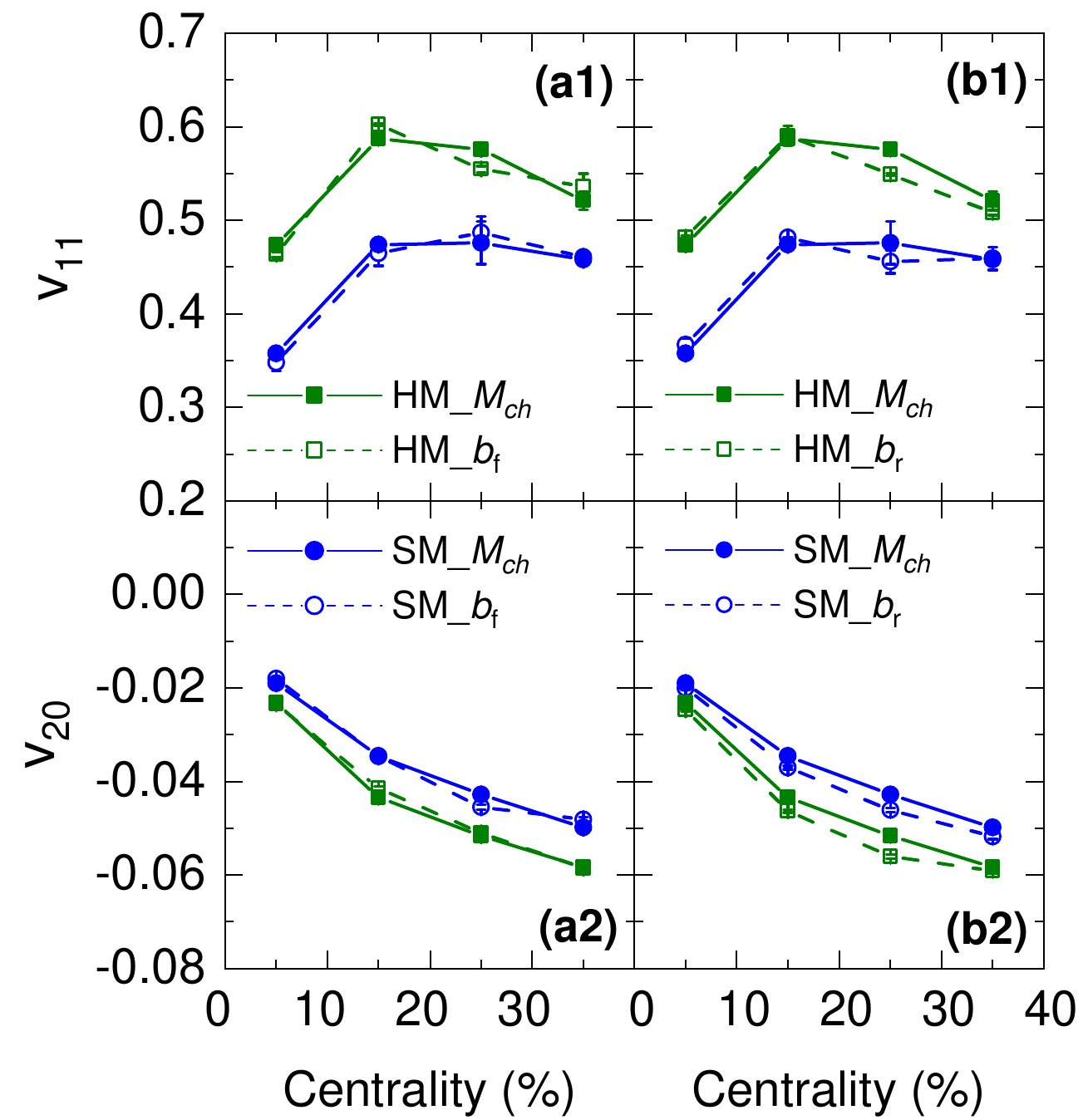}
    \caption{(Color online) Centrality dependence of directed flow slope $v_{11}$ [top panels (a1) and (b1)] and elliptic flow $v_{20}$ [bottom panels (a2) and (b2)] of free protons at mid-rapidity for four scenarios. Where error bars are not shown, they are smaller than the symbols.}
    \label{fig:6}
\end{figure}

To quantitatively evaluate the influence of EoS and centrality determination on the collective flow, Fig. \ref{fig:6} further shows the centrality dependence of the directed flow slope $v_{11}$ (top panels) and elliptic flow $v_{20}$ (bottom panels) of free protons at mid-rapidity with HM and SM, $M_\mathrm{ch}$, $b_{f}$ and $b_{r}$ filters. 
The $v_{11}$ and $v_{20}$ are extracted by assuming $v_1(y) = v_{11}y + v_{13}y^3$ and $v_2(y) = v_{20} + v_{22}y^2 + v_{24}y^4$ in the range of $|y| \textless 0.1$, and have been extensively used to constrain the nuclear EoS at different densities in HICs from intermediate to relativistic energies, for example Refs. \cite{Heinz:2013th,STAR:2014clz}. 
It can be found that the $v_{11}$ values obtained from event samples selected by different centrality determination methods are close to each other, and the values of HM are larger than those of SM. 
These are consistent with the fact that $v_{11}$ is insensitive to the centrality determination method but sensitive to the EoS, and a stiffer EoS will lead to a much stronger repulsive effect at the investigated energy \cite{Andronic:2006ra}. 
As for the $v_{20}$, the $|v_{20}|$ values of HM are also obviously larger than those with SM, and the values of the events selected by $b_r$ are slightly larger than those by $M_\mathrm{ch}$. 
From these results of collective flow, one can infer that the directed and elliptic flow parameters are more sensitive to the stiffness of the EoS than to the centrality determination method, but the inaccurate definition of centrality will have a certain impact on the elliptic flow parameter.

\subsection{Transverse Momentum Distribution}

\begin{figure}[t]
    \centering
    \includegraphics[width=0.49\textwidth]{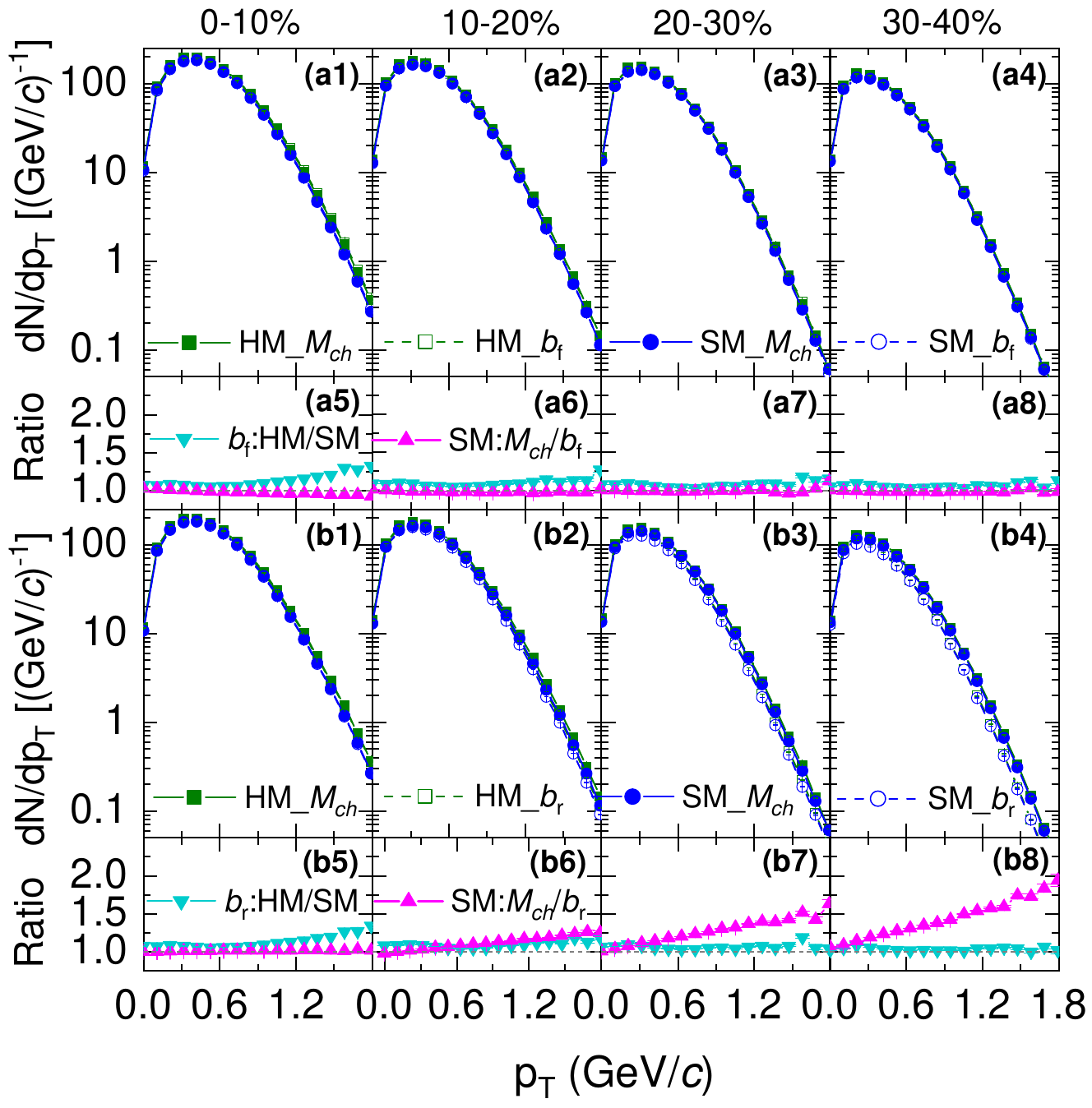}
    \caption{(Color online) The transverse momentum distribution [panels (a1)-(a4) and (b1)-(b4)] and the ratio [panels (a5)-(a8) and (b5)-(b8)] of free protons at centralities of 0-10\%, 10-20\%, 20-30\% and 30-40\%, respectively. The four scenarios are shown with different colors and line types same as in Fig. \ref{fig:4}.}
    \label{fig:7}
\end{figure}

Furthermore, since the transverse-momentum ($ p_\mathrm{T}$) distribution is closely related to the EoS and to the collective flow, 
the top panels of Fig. \ref{fig:7} depict the $p_\mathrm{T}$ distributions of free protons for various cases in 0-10\%, 10-20\%, 20-30\% and 30-40\% central collisions. 
And the ratio of these $p_\mathrm{T}$ distributions between SM and HM, $M_\mathrm{ch}$ and $b_{f}$ ($b_{r}$) are depicted by cyan down-triangles and magenta up-triangles in the corresponding bottom panels, respectively. 
It is seen that the results of HM are slightly higher than those of SM, and from left to right panels, the ratio $b_f$:HM/SM and $b_{r}$:HM/SM decrease with increasing centrality, which means as the centrality increases, the effect of EoS on free proton yield and $p_\mathrm{T}$ distribution decreases. 
In addition, the results of events selected by $M_\mathrm{ch}$ are similar to those of $b_{f}$ under the same EoS condition, the ratio SM:$M_\mathrm{ch}/b_{f}$ is basically unchanged with increasing centrality. 
By contrast, the results of events selected by $M_\mathrm{ch}$ are higher than those of $b_{r}$, and the ratio SM:$M_\mathrm{ch}/b_{r}$ increases with increasing centrality, which means the effect of the centrality determination method increases, particularly for higher $p_\mathrm{T}$, these results are consistent with Fig. \ref{fig:4}(b2). 
Since the density and the collision number per nucleon experienced will be decreased with the increase in centrality, and the protons with higher $p_\mathrm{T}$ tend to be emitted earlier, thus they are dominantly influenced by the initial geometry.

\begin{figure}[t]
    \centering
    \includegraphics[width=0.49\textwidth]{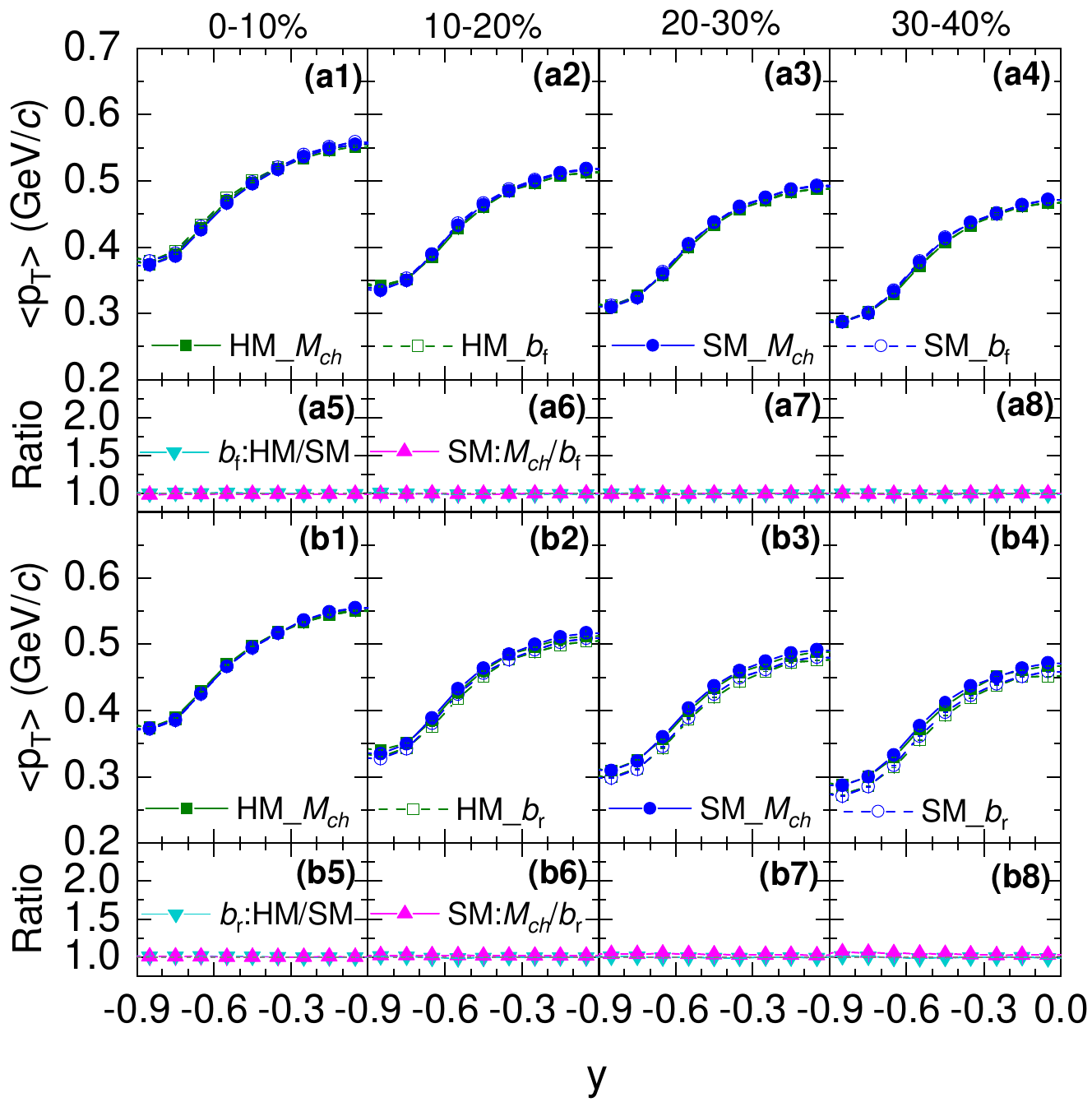}
    \caption{(Color online) Same as Fig. \ref{fig:7}, but for the rapidity dependence of the mean transverse momentum $\langle p_\mathrm{T} \rangle$.}
    \label{fig:8}
\end{figure}

In addition, Fig. \ref{fig:8} presents the rapidity $y$ dependent mean transverse momentum $\langle p_\mathrm{T} \rangle$ for various cases and the corresponding ratios for 0-10\%, 10-20\%, 20-30\% and 30-40\% central collisions. 
In all centrality cases, the results of SM and HM EoS under the same centrality definition are almost the same, and the results of $M_\mathrm{ch}$ and $b_{f}$ ($b_{r}$) under the same stiffness of EoS are also well consistent. 
Comparing with results shown in Figs. \ref{fig:4} and \ref{fig:7}, one can conclude that the $y$-dependent $\langle p_\mathrm{T} \rangle$ of free protons is insensitive to the effects of the EoS and the centrality determination method, while the yields and transverse momentum distributions are more sensitive to both two factors.

\section{Summary and outlook}\label{summ}

In this work, the effects of the nuclear EoS and the centrality determination method on nucleonic observables in Au+Au collisions at $\sqrt{s_{NN}}$ = 2.4 GeV are systematically investigated within the Ultra-relativistic Quantum Molecular Dynamics (UrQMD) framework. 
Simulations are performed using a soft and a hard momentum-dependent EoS, and event centrality is defined using three commonly employed methods, the charged-particle multiplicity $M_\mathrm{ch}$, a geometrical interpretation based on the range of input impact parameter $b_{f}$, and a Glauber Monte Carlo-based impact parameter selection $b_{r}$. 
The analysis demonstrates that event samples selected by different centrality methods exhibit significant differences in the real impact parameter distributions, with the most pronounced discrepancy observed between events selected using $M_\mathrm{ch}$ and $b_{r}$. These differences propagate into final-state observables and lead to non-negligible systematic effects. 
In particular, the free proton yield and transverse momentum distribution show a strong dependence on the choice of centrality definition, especially in peripheral collisions, whereas the directed flow slope $v_{11}$ and elliptic flow $v_{20}$ at mid-rapidity are primarily governed by the stiffness of the EoS and display a comparatively weaker sensitivity to centrality-related uncertainties. 

A key finding of this study is that when the geometrical-based centrality determination method ($b_{f}$) is employed, the uncertainties associated with centrality selection generally have a smaller effect on the yield, collective flow, transverse momentum distribution, and the mean transverse momentum of free protons than the effects induced by the EoS itself. 
In contrast, when the Glauber MC-based method ($b_{r}$) is used, the influence of centrality-related uncertainties becomes comparable to or even exceeds that of the EoS. This behavior indicates that some underlying assumptions of the Glauber MC framework, such as the proportionality between particle multiplicity and the number of participants, are not fully justified at low and intermediate collision energies, where nucleon-nucleon elastic cross section plays a more prominent role.

These results highlight the necessity of establishing a rigorous and self-consistent mapping between charged-particle multiplicity and collision geometry before final-state observables are used to impose quantitative constraints on the high-density nuclear EoS. Our study shows that, at SIS18 energies, the geometrical interpretation of centrality remains valid and is consistent with dynamical multiplicity-based selection, whereas the applicability of Glauber MC-based centrality determination method becomes increasingly limited in this energy regime.

As an outlook, future work will focus on employing Bayesian inference techniques to quantitatively assess the systematic deviations between theoretical simulations and experimental measurements, with particular emphasis on their effects on fluctuations and correlations across different centrality classes, rapidity, and transverse-momentum intervals. In addition, a more detailed investigation of the energy dependence and model assumptions underlying Glauber MC-based centrality determination method, especially its limitations at low and intermediate energies, will be essential for improving the reliability of EoS studies in this regime.

\begin{acknowledgments}
The authors acknowledge support by the computing server C3S2 Huzhou University. 
This work is supported in part by the National Natural Science Foundation of China (Nos. 12335008 and 12505143), 
the National Key Research and Development Program of China (No. 2023YFA1606402), 
the Zhejiang Provincial Natural Science Foundation of China (No. LQN25A050003), 
the Huzhou Natural Science Foundation (No. 2024YZ28), 
the Scientific Research Fund of the Zhejiang Provincial Education Department (No. Y202353782),
the Foundation of National Key Laboratory of Plasma Physics (Grant No. 6142A04230203),
the Fund for Shanxi "1331 Project" Key Subjects Construction.

\end{acknowledgments}


\begin{thebibliography}{99}
\bibitem{Danielewicz:2002pu}
P.~Danielewicz, R.~Lacey and W.~G.~Lynch,
Science \textbf{298}, 1592-1596 (2002).

\bibitem{Gyulassy:2004zy}
M.~Gyulassy and L.~McLerran,
Nucl. Phys. A \textbf{750}, 30-63 (2005).

\bibitem{Fukushima:2010bq}
K.~Fukushima and T.~Hatsuda,
Rept. Prog. Phys. \textbf{74}, 014001 (2011).

\bibitem{Andronic:2017pug}
A.~Andronic, P.~Braun-Munzinger, K.~Redlich and J.~Stachel,
Nature \textbf{561}, 321-330 (2018).



\bibitem{Frankfurt:1981mk}
L.~L.~Frankfurt and M.~I.~Strikman,
Phys. Rept. \textbf{76}, 215-347 (1981).

\bibitem{Jacobs:2004qv}
P.~Jacobs and X.~N.~Wang,
Prog. Part. Nucl. Phys. \textbf{54}, 443-534 (2005).


\bibitem{HADES:2017def}
J.~Adamczewski-Musch \textit{et al.} [HADES Collaboration],
Eur. Phys. J. A \textbf{54}, 85 (2018).

\bibitem{STAR:2022etb}
M.~Abdallah \textit{et al.} [STAR Collaboration],
Phys. Rev. C \textbf{107}, 024908 (2023).

\bibitem{Zhang:2017xda}
S.~Zhang, Y.~G.~Ma, J.~H.~Chen, W.~B.~He and C.~Zhong,
Phys. Rev. C \textbf{95}, 064904 (2017).

\bibitem{Cavata:1990gk}
C.~Cavata, M.~Demoulins, J.~Gosset \textit{et al.}
Phys. Rev. C \textbf{42}, 1760-1763 (1990).

\bibitem{Kim:1992zza}
Y.~d.~Kim, R.~T.~de Souza, D.~R.~Bowman \textit{et al.}
Phys. Rev. C \textbf{45}, 338-352 (1992).

\bibitem{FOPI:1996cjz}
W.~Reisdorf \textit{et al.} [FOPI Collaboration],
Nucl. Phys. A \textbf{612}, 493-556 (1997).


\bibitem{Li:2020lup}
P.~Li, Y.~Wang, Q.~Li, J.~Wang and H.~Zhang,
J. Phys. G \textbf{47}, 035108 (2020).

\bibitem{Li:2020qqn}
F.~Li, Y.~Wang, H.~L\"u \textit{et al.}
J. Phys. G \textbf{47}, 115104 (2020).

\bibitem{Li:2021plq}
F.~Li, Y.~Wang, Z.~Gao \textit{et al.}
Phys. Rev. C \textbf{104}, 034608 (2021).

\bibitem{Tsang:2021rku}
C.~Y.~Tsang, Y.~Wang, M.~B.~Tsang \textit{et al.}
[arXiv:2107.13985].

\bibitem{Mallick:2021wop}
N.~Mallick, S.~Tripathy, A.~N.~Mishra, S.~Deb and R.~Sahoo,
Phys. Rev. D \textbf{103}, 094031 (2021).

\bibitem{Sorensen:2023zkk}
A.~Sorensen, K.~Agarwal, K.~W.~Brown, Z.~Chaj\k{e}cki, P.~Danielewicz, C.~Drischler, S.~Gandolfi, J.~W.~Holt, M.~Kaminski and C.~M.~Ko, \textit{et al.}
Prog. Part. Nucl. Phys. \textbf{134}, 104080 (2024).

\bibitem{TMEP:2022xjg}
H.~Wolter \textit{et al.} [TMEP Collaboration],
Prog. Part. Nucl. Phys. \textbf{125}, 103962 (2022).

\bibitem{Fuchs:2000kp}
C.~Fuchs, A.~Faessler, E.~Zabrodin and Y.~M.~Zheng,
Phys. Rev. Lett. \textbf{86}, 1974-1977 (2001).

\bibitem{Roca-Maza:2018ujj}
X.~Roca-Maza and N.~Paar,
Prog. Part. Nucl. Phys. \textbf{101}, 96-176 (2018).

\bibitem{Huth:2021bsp}
S.~Huth, P.~T.~H.~Pang, I.~Tews \textit{et al.}
Nature \textbf{606}, 276-280 (2022).

\bibitem{Tsang:2023vhh}
C.~Y.~Tsang, M.~B.~Tsang, W.~G.~Lynch, R.~Kumar and C.~J.~Horowitz,
Nature Astron. \textbf{8}, 328-336 (2024).

\bibitem{STAR:2021ozh}
M.~S.~Abdallah \textit{et al.} [STAR Collaboration],
Phys. Lett. B \textbf{827}, 136941 (2022).

\bibitem{Li:2022iil}
P.~Li, J.~Steinheimer, T.~Reichert, A.~Kittiratpattana, M.~Bleicher and Q.~Li,
Sci. China Phys. Mech. Astron. \textbf{66}, 232011 (2023).

\bibitem{Li:2025iqq}
B.~A.~Li and W.~J.~Xie,
Phys. Rev. C \textbf{111}, 054602 (2025).


\bibitem{Oertel:2016bki}
M.~Oertel, M.~Hempel, T.~Kl\"ahn and S.~Typel,
Rev. Mod. Phys. \textbf{89}, 015007 (2017).

\bibitem{Li:2024lmd}
J.~J.~Li, A.~Sedrakian and M.~Alford,
Astrophys. J. \textbf{967}, 116 (2024).

\bibitem{Ye:2024meg}
J.~T.~Ye, R.~Wang, S.~P.~Wang and L.~W.~Chen,
Astrophys. J. \textbf{985}, 238 (2025).

\bibitem{OmanaKuttan:2023cno}
M.~Omana Kuttan, J.~Steinheimer, K.~Zhou, M.~Bleicher and H.~Stoecker,
Eur. Phys. J. C \textbf{83}, 792 (2023).


\bibitem{Bleicher:2022kcu}
M.~Bleicher and E.~Bratkovskaya,
Prog. Part. Nucl. Phys. \textbf{122}, 103920 (2022).

\bibitem{Bass:1998ca}
S.~A.~Bass, M.~Belkacem, M.~Bleicher \textit{et al.}
Prog. Part. Nucl. Phys. \textbf{41}, 255-369 (1998).

\bibitem{Bleicher:1999xi}
M.~Bleicher, E.~Zabrodin, C.~Spieles \textit{et al.}
J. Phys. G \textbf{25}, 1859-1896 (1999).

\bibitem{Steinheimer:2018rnd}
J.~Steinheimer, Y.~Wang, A.~Mukherjee, Y.~Ye, C.~Guo, Q.~Li and H.~Stoecker,
Phys. Lett. B \textbf{785}, 40-45 (2018).

\bibitem{Li:2021sdc}
P.~Li, Y.~Wang, J.~Steinheimer, Q.~Li and H.~Zhang,
Phys. Lett. B \textbf{818}, 136393 (2021).

\bibitem{Wang:2020vwb}
Y.~J.~Wang and Q.~F.~Li,
Front. Phys. (Beijing) \textbf{15}, 44302 (2020).

\bibitem{Li:2011zzp}
Q.~Li, C.~Shen, C.~Guo, Y.~Wang, Z.~Li, J.~Lukasik and W.~Trautmann,
Phys. Rev. C \textbf{83}, 044617 (2011).


\bibitem{Ye:2018vbc}
Y.~Ye, Y.~Wang, J.~Steinheimer, Y.~Nara, H.~Xu, P.~Li, D.~Lu, Q.~Li and H.~Stoecker,
Phys. Rev. C \textbf{98}, 054620 (2018).

\bibitem{Du:2023ype}
H.~Du, G.~F.~Wei and G.~C.~Yong,
Phys. Lett. B \textbf{839}, 137823 (2023).

\bibitem{Steinheimer:2024eha}
J.~Steinheimer, T.~Reichert, Y.~Nara and M.~Bleicher,
J. Phys. G \textbf{52}, 035103 (2025).

\bibitem{Wang:2013wca}
Y.~Wang, C.~Guo, Q.~Li, H.~Zhang, Z.~Li and W.~Trautmann,
Phys. Rev. C \textbf{89}, 034606 (2014).



\bibitem{Zhang:2012qm}
Y.~Zhang, Z.~Li, C.~Zhou and M.~B.~Tsang,
Phys. Rev. C \textbf{85}, 051602 (2012).

\bibitem{Li:2016wkb}
Q.~Li, Y.~Wang, X.~Wang and C.~Shen,
Sci. China Phys. Mech. Astron. \textbf{59}, 672013 (2016).

\bibitem{Kireyeu:2022qmv}
V.~Kireyeu, J.~Steinheimer, J.~Aichelin, M.~Bleicher and E.~Bratkovskaya,
Phys. Rev. C \textbf{105}, 044909 (2022).


\bibitem{Li:2017tom}
L.~Li, Y.~Zhang, Z.~Li, N.~Wang, Y.~Cui and J.~Winkelbauer,
Phys. Rev. C \textbf{97}, 044606 (2018).


\bibitem{JiaYe:2021lgg}
J. Y. Yang, L.~Liu, P. C. Li, Y. J. Wang and Q. F. Li,
Sci. Sin. Phys. Mech. Astro. \textbf{51}, 112011 (2021).

\bibitem{FOPI:2010xrt}
W.~Reisdorf \textit{et al.} [FOPI Collaboration],
Nucl. Phys. A \textbf{848}, 366-427 (2010).

\bibitem{Li:2018wpv}
P.~Li, Y.~Wang, Q.~Li, C.~Guo and H.~Zhang,
Phys. Rev. C \textbf{97}, 044620 (2018).

\bibitem{Xiao:2023pqs}
K.~Xiao, P.~Li, Y.~Wang, F.~Liu and Q.~Li,
Nucl. Sci. Tech. \textbf{34}, 62 (2023).

\bibitem{Stoecker:1986ci}
H.~Stoecker and W.~Greiner,
Phys. Rept. \textbf{137}, 277-392 (1986).

\bibitem{Reichert:2023eev}
T.~Reichert, O.~Savchuk, A.~Kittiratpattana, P.~Li, J.~Steinheimer, M.~Gorenstein and M.~Bleicher,
Phys. Lett. B \textbf{841}, 137947 (2023).

\bibitem{Reisdorf:1997fx}
W.~Reisdorf and H.~G.~Ritter,
Ann. Rev. Nucl. Part. Sci. \textbf{47}, 663-709 (1997)

\bibitem{Herrmann:1999wu}
N.~Herrmann, J.~P.~Wessels and T.~Wienold,
Ann. Rev. Nucl. Part. Sci. \textbf{49}, 581-632 (1999).

\bibitem{Reichert:2024ayg}
T.~Reichert and J.~Aichelin,
Phys. Rev. C \textbf{111}, 054916 (2025).

\bibitem{Snellings:2011sz}
R.~Snellings,
New J. Phys. \textbf{13}, 055008 (2011).

\bibitem{Gao:2022shr}
B.~Gao, Y.~Wang, Z.~Gao and Q.~Li,
Phys. Lett. B \textbf{838}, 137685 (2023).

\bibitem{Wang:2024ktk}
Y.~Wang, B.~Gao, G.~Wei, P.~Li and Q.~Li,
Phys. Rev. C \textbf{110}, 044606 (2024).

\bibitem{Voloshin:2011mx}
S.~A.~Voloshin,
Prog. Part. Nucl. Phys. \textbf{67}, 541-546 (2012).

\bibitem{Heinz:2013th}
U.~Heinz and R.~Snellings,
Ann. Rev. Nucl. Part. Sci. \textbf{63}, 123-151 (2013).

\bibitem{HADES:2020lob}
J.~Adamczewski-Musch \textit{et al.} [HADES Collaboration],
Phys. Rev. Lett. \textbf{125}, 262301 (2020).


\bibitem{Steinheimer:2022gqb}
J.~Steinheimer, A.~Motornenko, A.~Sorensen, Y.~Nara, V.~Koch and M.~Bleicher,
Eur. Phys. J. C \textbf{82}, no.10, 911 (2022)

\bibitem{Tarasovicova:2024isp}
L.~A.~Tarasovi{\v{c}}ov{\'a}, J.~Mohs, A.~Andronic, H.~Elfner and K.~H.~Kampert,
Eur. Phys. J. A \textbf{60}, no.11, 232 (2024)

\bibitem{HADES:dndy}
M. Szala,
Light nuclei formation in heavy ion collisions measured with HADES,
report, 
https://indico.ectstar.eu/event/52/contributions


\bibitem{FOPI:2011aa}
W.~Reisdorf \textit{et al.} [FOPI Collaboration],
Nucl. Phys. A \textbf{876}, 1-60 (2012).


\bibitem{STAR:2014clz}
L.~Adamczyk \textit{et al.} [STAR Collaboration],
Phys. Rev. Lett. \textbf{112}, 162301 (2014).


\bibitem{Andronic:2006ra}
A.~Andronic, J.~Lukasik, W.~Reisdorf and W.~Trautmann,
Eur. Phys. J. A \textbf{30}, 31-46 (2006)




\end{thebibliography}
\end{document}